\documentclass[aps,pra,a4paper,superscriptaddress,numbers,longbibliography,showpacs,showkeys,floatfix,svgnames,reprint,nofootinbib]{revtex4-2}

\usepackage{amsfonts,amssymb,amsmath}
\usepackage{hyperref,xcolor,graphicx}
\usepackage[separate-uncertainty=true]{siunitx}
\usepackage[ISO]{diffcoeff}

\DeclareSIUnit\gauss{G}

\usepackage[english]{babel}
\usepackage[utf8]{inputenc}
\usepackage[T1]{fontenc}

\usepackage{xspace}
\usepackage{mathtools}
\usepackage{physics}
\usepackage{braket}

\newcommand{\figref}[2]{\hyperref[#1]{\ref{#1}{(#2)}}}
\addto\captionsenglish{}

\newcommand{\NA}{\ensuremath{\mathrm{NA}}} \newcommand{\zR}{\ensuremath{z_\mathrm{R}}} \newcommand{\dz}{\ensuremath{d_\mathrm{L}}} \newcommand{\vect}[1]{\ensuremath{\boldsymbol{#1}}} \renewcommand{\i}{\mathrm i} \newcommand{\e}{\mathrm e} \newcommand{\cvacuum}{\ensuremath{c}} \renewcommand{\Re}{\operatorname{Re}}  

\def\theTitle{%
Three-dimensional imaging of single atoms in an optical lattice via helical point-spread-function engineering
}

\hypersetup{
 pdftitle = {\theTitle},
 pdfauthor = {Tangi Legrand, Falk-Richard G. Winkelmann, Wolfgang Alt, Dieter Meschede, Andrea Alberti, Carrie A. Weidner},
	colorlinks=true,linkcolor=blue,citecolor=blue,filecolor=blue,urlcolor=blue,pdfpagemode=UseNone,pdfstartview={XYZ null null 1.00}
}

\renewcommand\textemdash{\leavevmode\unskip\kern0.8pt---\kern1pt\ignorespaces}

\begin{document}
\title{\theTitle}

\author{Tangi Legrand}
\affiliation{Institut für Angewandte Physik, Universität Bonn, Wegelerstraße 8, D-53115 Bonn, Germany}
\author{Falk-Richard Winkelmann}
\affiliation{Institut für Angewandte Physik, Universität Bonn, Wegelerstraße 8, D-53115 Bonn, Germany}
\author{Wolfgang Alt}
\affiliation{Institut für Angewandte Physik, Universität Bonn, Wegelerstraße 8, D-53115 Bonn, Germany}
\author{Dieter Meschede}
\affiliation{Institut für Angewandte Physik, Universität Bonn, Wegelerstraße 8, D-53115 Bonn, Germany}
\author{Andrea Alberti}
 \email{alberti@iap.uni-bonn.de}
\affiliation{Institut für Angewandte Physik, Universität Bonn, Wegelerstraße 8, D-53115 Bonn, Germany}
\author{Carrie A. Weidner}
 \email{c.weidner@bristol.ac.uk}
\affiliation{Quantum Engineering Technology Laboratories, H. H. Wills Physics Laboratory and Department of Electrical, Electronic, and Mechanical Engineering, University of Bristol, Bristol BS8 1FD, UK}

\date{\today}

\begin{abstract}
	We demonstrate a method for determining the three-dimensional location of single atoms in a quantum gas microscopy system using a phase-only spatial light modulator to modify the point-spread function of the high-resolution imaging system. Here, the typical diffracted spot generated by a single atom as a point source is modified to a double spot that rotates as a function of the atom's distance from the focal plane of the imaging system. We present and numerically validate a simple model linking the rotation angle of the point-spread function with the distance to the focal plane. 
    We show that, when aberrations in the system are carefully calibrated and compensated for, this method can be used to determine an atom's position to within a single lattice site in a single experimental image, extending quantum simulation with microscopy systems further into the regime of three dimensions.
\end{abstract}

\maketitle

\section{\label{sec:introduction}Introduction}

    The advent of quantum gas microscopy, wherein one can manipulate and image atoms with high resolution, has revolutionized the field of analog quantum simulation. The first quantum gas microscopes were based on fluorescence imaging of ultracold bosonic atoms trapped in cubic optical lattices placed near a high-resolution microscope objective~\cite{greiner2009, kuhr2010}, and in recent years, fermionic systems~\cite{thywissen2015, kuhr2015, greiner2015, gross2015, zwierlein2015}, molecular systems~\cite{Bakr2022} and systems with non-cubic lattice shapes~\cite{schauss2021, fukuhara2020} have been demonstrated. These systems have been used to study, among others, entanglement dynamics~\cite{greiner2015a}, many-body localization~\cite{gross2019, greiner2019}, and the Fermi-Hubbard model~\cite{greif2019, zwierlein2020, bloch2021}.

    However, these lattice-based systems have largely been limited to studies of quantum simulation in two dimensions, where a single plane of the lattice has been prepared, with some exceptions for bilayer studies~\cite{greiner2015c, bloch2020a}. This is, in no small part, due to the fact that, like with conventional microscopy systems, the imaging signal is obtained by integrating the atomic fluorescence along the direction of the optical axis of the microscope. 
    Additionally, given a selected lattice plane, it is extremely difficult to distinguish the fluorescence signal emitted by in-plane atoms from that of out-of-plane atoms, since the diffraction-limited depth of focus of the quantum gas microscope can extend over several planes, even at high numerical apertures $\mathrm{NA}\approx 0.6$. Tomographic methods, where multiple images are taken while the microscope objective is moved in between exposures, were first demonstrated for large-spacing lattices~\cite{weiss2007, schmiedmayer2009} and later for lattices with half-wavelength plane spacing~\cite{sherson2020}, but these methods' multiple exposure requirement limits their efficacy due to long imaging times.

    Pulsed-ion microscopes, wherein the atoms are ionized based on their position, have been demonstrated~\cite{pfau2021}, but detection efficiencies are low~\cite{ott2016} and they are limited to $\approx \SI{1}{\micro\meter}$ resolution, thus rendering them unable to distinguish between adjacent planes of a typical cubic lattice with $\approx\SI{500}{\nano\meter}$ plane spacing. Moreover, such ionizing techniques do not allow one to reuse the same atoms for subsequent measurements. 
    Three-dimensional imaging methods with micro-lens arrays~\cite{burke2017} and holographic imaging methods~\cite{ahn2021} also suffer from low spatial resolution, although a recent holographic method--while still requiring multiple exposures--overcomes the conventional axial resolution limitations and could be applied to quickly image a bulk cloud in three dimensions~\cite{wu2023}. 
    However, a high-resolution imaging technique based on coupling atomic fluorescence into a multimode fibre (instead of a high-resolution objective) has shown promise for three-dimensional reconstruction of the atoms' positions~\cite{ourjoumtsev2021}.

    In this work, we demonstrate three-dimensional imaging of single atoms in a quantum gas microscope using point-spread-function (PSF) engineering of the atoms' fluorescence signal. Inspired by similar work in the imaging of bioluminescent molecules~\cite{moerner2009}, we modify the phase of the atoms' fluorescence with a spatial light modulator (SLM) placed in the Fourier plane of the imaging system. In this way, we encode each atom's position along the optical axis of the imaging system (here, the $z$ direction). Specifically, the typical point-like PSF of a single atom is modified into a two-point, double-helix PSF (DH-PSF) that rigidly rotates as a function of its $z$-position.
    This is done by writing the phase of a superposition of Laguerre-Gaussian modes onto the SLM~\cite{shamir1996, shamir2000, piestun2006}. In particular, we show that this method has promise with regards to determining the three-dimensional distribution of atoms in a lattice system, when aberrations in the system are carefully controlled.

    This paper is organized as follows. In Sec.~\ref{sec:theory} we describe the basic theoretical aspects of the rotating PSF. We describe the experimental setup in Sec.~\ref{sec:expt} and present our experimental results in Sec.~\ref{sec:results}. Simulations of the effects of aberrations in the experimental system are presented in Sec.~\ref{sec:sims}. Sec.~\ref{sec:conc} discusses our results and concludes.

\section{\label{sec:theory}Theory}

    The theoretical background for rigidly rotating PSFs has been discussed in detail in Refs.~\cite{piestun2006, moerner2009, shamir1996, shamir2000}, and we present only the most relevant aspects here. 
    In general, in the paraxial approximation, one can construct a rotating PSF from superpositions of Laguerre-Gaussian modes~\cite{shamir1996}, which are themselves a complete basis set of solutions to the paraxial Helmholtz equation in cylindrical coordinates. 
    We denote these modes as $\ket{l,p}$, and they can be written spatially as~\cite{shamir2000}
    \begin{multline}
        \label{eq:LG}
        \braket{\vect{x}|l,p} = C_{lp} \,
        \frac{w_0}{w(\tilde{z})}\left(\sqrt{2}\,\tilde{r}\right)^{|l|}
			\exp\left(-\tilde{r}^2\right)
			L_p^{|l|}\left(2\tilde{r}^2\right) \\
			\times \exp\left( \i \tilde{r}^2\tilde{z}\right)
			\exp\left(\i l\phi -\i\psi_{lp}(\tilde{z})\right)\,,
    \end{multline}
    where $\tilde{z} = z/\zR$ is the longitudinal coordinate (along the optical axis) scaled by the Rayleigh length $\zR = \pi w_0^2/\lambda$ for a Gaussian spot size $w_0$ and light wavelength $\lambda$. 
    The Gaussian beam radius changes as a function of $\tilde{z}$ as $w(\tilde{z}) = w_0\sqrt{1 + \tilde{z}^2}$. 
    The scaled radial coordinate is  $\tilde{r} = r/w(\tilde{z})$. 
    Note that in generating the phase fronts, we work at $z = 0$ such that $w(\tilde{z}) = w_0$. 
    The azimuthal coordinate is denoted by $\phi\in[0, 2\pi)$. 
    The variable $C_{lp}$ is a normalization factor given by 
    \begin{equation}
        C_{lp} = \sqrt{\frac{2p!}{\pi(p+\abs{l})!}}\,.
    \end{equation}
    The functions making up the orthogonal Laguerre-Gaussian basis are given by the azimuthal mode number $l\in\mathbb{Z}$, the radial mode number $p\in \mathbb{N}$ and the combined mode number $n=2p+\abs{l}$.
    The $L^l_p$ denote the generalized Laguerre polynomials, and $\psi_{lp}(\tilde{z}) = (n+1)\arctan{(\tilde{z})}$ is the Gouy phase, which is zero at $z = 0$.
    As $n$ increases, the spatial extent of the beam becomes larger. 
    The value of $l$ denotes the number and handedness of the azimuthal phase windings and there are $p$ radial zeros for each mode. 

    Each of these Laguerre-Gaussian modes is an eigenmode of rotation about the $z$ axis, meaning that the intensity distribution of the light $I\propto |\braket{\vect{r}|l,p}|^2$ is axially symmetric and stationary. 
    Thus, in order to make a PSF rotate, we must superpose two or more of these modes. 
    In general, one can show that, given a superposition of modes $\{\ket{l_j,p_j}\}$ such that $\Delta n = n_{j+1} - n_j$ and $\Delta l = l_{j+1}-l_j$ are equal for all $j$ (ordered sequentially for increasing $n$), the interference terms in the superposition rotate about the optical axis at the rate
    \begin{equation}
        \label{eq:phi_rot}
    \diff{\phi}{z} = V\diff{}{z}\arctan{(z/\zR)},
    \end{equation}
    where $V = \Delta n/\Delta l$. 
    Here, and moving forward, we return to using $z/\zR$ over $\tilde{z}$ for clarity.
    Combinations of modes satisfying the above condition lie along a line in the Laguerre-Gaussian modal plane~\cite{piestun2008a}. 
    The superposition has an intensity distribution whose axial rotational symmetry is broken due to the interference of modes of different orbital angular momentum phases $\e^{\i l_j\phi}$ (for $\Delta l\ne 0$).
    The dynamics along the optical axis are provided by the different Gouy phases $\psi_{l_jp_j}(z)$ (for $\Delta n\ne 0$).
    Overall, the interference pattern performs a rigid axial rotation along $z$ (scaled by $w(z)$).
    Details are given in Appendix~\ref{app:lgsuperpositions}.
    The total rotation from $z = -\infty$ to $z = \infty$ is given by $V\pi$, so beams with higher $\Delta n$ will have larger spatial extents but will also rotate faster. 
    Half of this rotation occurs within $[-\zR, \zR]$; outside the Rayleigh range, the rotating rate decreases.
    For a given distance $z$ away from the focal plane, we get from Eq.~\eqref{eq:phi_rot} the rotation angle of the PSF
    \begin{equation}
        \label{eq:rotationangle}
        \theta = V\arctan{(z/\zR)} + \alpha\,,
    \end{equation}
    where $\alpha$ is the (arbitrary) measured angle of the rotating PSF in the focal plane.

\begin{figure}[t]
        \centering
        \includegraphics[width=\columnwidth]{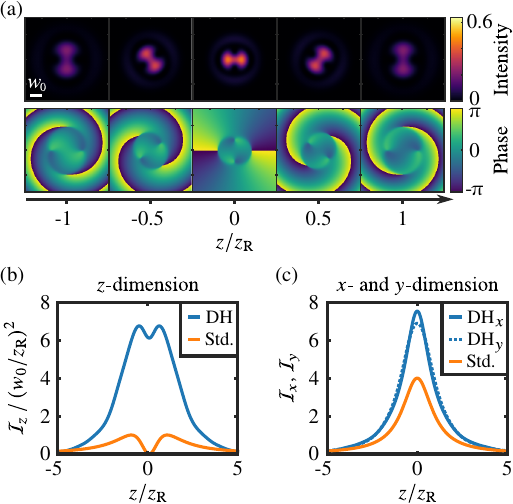}
        \caption{%
            (a)~Intensity and phase profile of the DH-PSF at different axial positions within the Rayleigh range. The upper end of the intensity color scale corresponds to the maximum intensity of the unmodified PSF. \mbox{(b)-(c)}~Fisher information $\mathcal{I}_\eta$ as a function of axial position $z/\zR$ of the DH-PSF (blue) and the standard PSF (Std., orange) with respect to the (b)~axial z-dimension, (c)~x-dimension (solid) and y-dimension (dashed). Note the different units of the ordinates, as well as the fact that $\mathcal{I}_z$ = 0 at $z = 0$ for the standard PSF.
        }
        \label{fig:dhpsf_fisherinformation}
    \end{figure}

    As an example PSF, consider the superposition $\ket{l,p}=\left(\ket{0,0}+\ket{2,1}\right)/\sqrt{2}$, whose transverse intensity and phase are shown in Fig.~\ref{fig:dhpsf_fisherinformation}~(a).
    This mode combination has a double-helical structure, hence the name double-helix PSF. 
    With $V=2$ the rotation angle is unambiguous within the Rayleigh range. 
    The topology of the intensity pattern is particularly well-suited for determining both the angle of rotation and the center position while leaving the energy as spatially focused as possible.
    This is also the reason why we use this particular PSF in this work.

    A further advantage provided by this DH-PSF is given by the degree to which one can localize an atom in three-dimensional space.
    For an atom positioned in $(x,y,z)$, the classical Fisher information informs us how efficient our estimate of its position can possibly be.
    It is defined as
    \begin{equation}
        \label{eq:fisherinformation}
        \mathcal{I}_\eta = \operatorname{E}\left[ \left.\left(\diffp{}{\eta}\ln f(X;\eta)\right)^2\,\right|\eta \right]\,,
    \end{equation}
    where $f(X;\eta)$ is the probability density function for an observable $X$, given a parameter $\eta=x,y$ or $z$.
    More precisely, the reciprocal of the Fisher information sets a lower bound for the variance of an unbiased estimator that uses the detected fluorescence signal to reconstruct the atom's position.
    Thus, the Fisher information is independent of how we construct the estimator based on the observed fluorescence signal, and it provides a measure of how efficient an (unbiased) estimator of the atom's position can be. Comparing the Fisher information of the DH-PSF with the ordinary PSF, we can directly quantify how much more information about the atom's position we can extract from the detected fluorescence signal.  
    By identifying the normalized transverse intensity distribution as a probability density, we compute---as a function of $z$---the Fisher information with respect to each Cartesian coordinate for the DH-PSF and, as comparison, for the fundamental Laguerre-Gaussian mode, as shown in Fig.~\ref{fig:dhpsf_fisherinformation}~(b) and~(c).
    The latter can be considered the standard, Airy-disk PSF in terms of Laguerre-Gaussian modes; this is what is commonly seen in quantum gas microscopy experiments. We do not assume any noise here. A complete noise model would allow us to quantify the resolution achievable in the experiment using the theoretical Fisher information, but this is outside of the scope of this work. Details on the computation can be found in Appendix~\ref{app:fisherinformation}.
    As can be seen, the DH-PSF provides a higher Fisher information for the localization along all dimensions as compared to using the standard PSF. 
    In particular, unlike the standard PSF, it provides high axial information even near the focus, as evident from Fig.~\ref{fig:dhpsf_fisherinformation}~(b). 
    Remarkably, the Fisher information of the standard PSF vanishes for $z = 0$, indicating its inability to differentiate between positive and negative positions of the atom along the z-axis, while the Fisher information of the DH-PSF is close to its maximum near the focus.
    The superior axial information of the double helix thus provides a decisive advantage over merely relying on the defocus.

    Regarding localization along the lateral dimensions, it is interesting to note that the DH-PSF yields larger Fisher information over the entire Rayleigh range, as can be seen in Fig.~\ref{fig:dhpsf_fisherinformation}~(c), which also improves the effective depth of field.
    However, this does not necessarily give the DH-PSF a better performance for applications with emitters in a single lateral plane, as the larger extent of the PSF makes the localization of closely spaced emitters less trivial, and the peak intensity is lower for a limited photon budget.
    At the focus, the DH-PSF consists of two lobes shifted from the origin along the $x$-axis. This azimuthal asymmetry results in a slightly higher maximal Fisher information along the $x$-dimension than the $y$-dimension.

    Finally, it is important to note that the PSFs based on Laguerre-Gaussian mode superpositions are not the only possible engineered PSFs that can be applied to a given system~\cite{moerner2014}, but these ensure rigid rotation about the focal plane of the imaging system, so they provide an excellent testbed for the proof-of-principle three-dimensional imaging demonstrated in this work.

\section{\label{sec:expt}Experimental setup}

    The experimental system is similar to that used in Refs.~\cite{alberti2016,alberti2021b}  and described in detail in Ref.~\cite{brakhane2017}. 
    In particular, we trap ultracold $^{133}$Cs atoms in a three-dimensional cubic lattice positioned near a custom \emph{in vacuo} high numerical aperture (NA) objective with a maximum NA of $0.92$~\cite{alberti2017} capable of site-resolved imaging of the individual atoms in the lattice. The horizontal lattice is created by a three-beam lattice with a wavelength of $\lambda_\mathrm{H} = \SI{866}{\nano\meter}$, giving rise to a single-site lattice spacing of $\lambda_\mathrm{H}/\sqrt{2} \approx \SI{612}{\nano\meter}$. The vertical lattice is provided by a retro-reflected lattice with wavelength $\lambda_\mathrm{V} = \SI{1064}{\nano\meter}$, so the distance between adjacent vertical lattice planes is $\dz=\SI{532}{\nano\meter}$; it is this vertical spacing that we seek to resolve in this work.

    The atoms are prepared first by standard cooling in a magneto-optical trap (MOT) and polarization gradient cooling, after which we load the atoms into the three-dimensional lattice. 
    In order to sparsely load the lattice, we hold the atoms in the lattice for $\SI{5}{\second}$ and resolve, on average $\num{6}$ atoms in the lattice when we begin imaging. It is important to note that no plane selection is performed, so we load many ($>10$) vertical lattice planes per experimental run.

    \begin{figure}[t]	
    	\centering
        \includegraphics[width=\columnwidth]{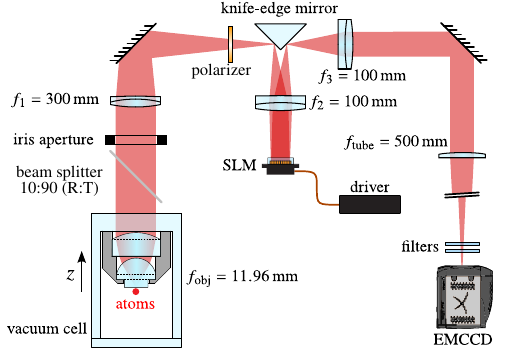}
    	\caption{A schematic of the fluorescence imaging system. The atoms are trapped in a three-dimensional lattice below a high-NA \emph{in vacuo} microscope objective. After passing through a $90/10$ beamsplitter (which allows for the vertical MOT beam to enter the chamber), the light is then re-imaged onto an EMCCD camera with a magnification of $M = \frac{f_1}{f_\mathrm{obj}}\frac{f_\mathrm{tube}}{f_3}$. In the Fourier plane of this re-imaging system, the phase of the atoms' fluorescence is modified by an SLM, which gives rise to the helically-diffracting PSF that encodes information on the atoms' $z$ position along the optical axis.
    	}
    	\label{fig:expt_schematic}
    \end{figure}

    The imaging system used in the experiment is shown in Fig.~\ref{fig:expt_schematic}. 
    The light collected by the high-NA objective is re-imaged onto an electron-multiplying (EM) CCD camera. 
    In an intermediate Fourier plane of this imaging system, we place an SLM (Santec SLM-100, with $1440\times1050$~pixels and a pixel pitch of $\SI{10.4}{\micro\meter}$) that modifies the phase of the imaging light with a modulation depth $>4\pi$, a $10$ bit resolution, and a typical response time of $\SI{100}{\milli\second}$.
    A polarizer placed before the SLM allows us to only select the light polarization that is modified by the SLM, and the use of a knife-edge mirror allows us to direct the light into the SLM at an angle near normal incidence, where the SLM is designed to operate. 
    The knife-edge mirror must be carefully aligned so that the beam is significantly narrower on its surface, but does not correspond exactly to the intermediate image planes, otherwise any defect on the mirror surface will be imaged directly onto the camera.
    We calibrate the SLM phase mask to account for curvature of the SLM chip as well as aberrations in the imaging setup.\footnote{This aberration correction is done empirically by adding Zernike polynomials to the SLM and optimizing the PSF of the atoms without the DH-PSF phase mask applied.} 
    The SLM also allows us to apply a holographic lens, enabling us to shift the focal plane of the imaging system by a given amount and thus focus on different vertical lattice planes (for details see Appendix~\ref{app:holographiclens}). 
    Nominally, the imaging system is designed to make full use of the $0.92$~NA microscope objective, but a motorized iris above the objective allows the effective NA of the imaging system to be reduced, which increases the depth of field of the system at the expense of horizontal resolution.

    The imaging system is designed as a 4f correlator system, and it allows a multiplication in Fourier space by a field whose Fourier transform gives the desired PSF.
    In image space, this is equivalent to convolving the lateral atom positions with the PSF.
    To create the DH-PSF with the desired Laguerre-Gaussian mode superposition in image space, we program the phase of this field onto the SLM placed in the pupil.
    This is possible because the Laguerre-Gaussian modes are eigenfunctions of the the two-dimensional Fourier transform expressed in polar coordinates~\cite{Pei:2012}.
    Due to the fact that the amplitude is not modulated in the pupil and due to diffraction at the finite aperture (which acts as a low-pass filter), the possible mode fidelity in image space is constrained.
    As reported in Ref.~\cite{hara2008}, achievable mode fidelities reach around $\SI{80}{\percent}$ at certain ratios of aperture radius and Laguerre-Gaussian mode waist $a/w_0$, with values around $a/w_0=\num{2}$ to $\num{4}$ for low mode numbers $p,l\le \num{5}$, depending on the specific mode.
    Here, $w_0$ is the waist of the phase mask in the pupil plane.
    In the case of our Laguerre-Gaussian mode superposition, we manually optimize $w_0$ in an optical test setup until the resulting PSF is as close as possible to the desired DH-PSF, resulting in $a/w_0=\num{2.97}$.
    As our simulations confirm (cf.\ Sec.~\ref{sec:sims}), the mode fidelity is sufficient, since the rotational properties do not deviate significantly from the simple model of scaled-rigid rotation Eq.~\eqref{eq:rotationangle}, except for very large NAs.

    We take three images of the atom cloud with the modified PSF via  fluorescence imaging on the D2 line of cesium ($\lambda=\SI{852}{\nano\meter}$). 
    For our intermediate NA of $\num{0.6}$, compared to the maximal NA, both the axial resolution through the standard PSF and the rotation angle per vertical lattice plane of the rotating PSF are smaller, thereby posing a greater challenge to vertical plane resolution than at higher NA.  
    In addition, due to lower photon collection efficiency, the reduced solid angle also results in a smaller signal-to-noise ratio.
    Finally, similar numerical aperture values are typical in ultracold atom experiments, where other factors such as the need for long working distance, large field of view, and chromatic corrections often prevent reaching values higher than $\NA=\num{0.6}-\num{0.7}$. 
    Thus, the demonstration of the use of the DH-PSF at the intermediate $\NA=\num{0.6}$ is of greater relevance.

    By using the SLM to vary the focal plane in between images (in addition to engineering the PSF and correcting for aberrations), the first and third images are set to image a focal plane at position $z_0=0$, while the second image focuses on a plane $z_0 + n \dz$, where $n$ is an integer that describes the number of vertical lattice planes by which we have shifted the focal plane. 
    In this work, we vary $n$ from $-4$ to $4$ in steps of $2$.
    Because our system was not optimized to eliminate atom hopping in the lattice (horizontally or vertically), we take the third image in order to post-select on atoms that have not hopped in any direction. 
    Our imaging exposure times were $\SI{1}{\second}$, with a delay of $\SI{500}{\milli\second}$ between images to allow the SLM to update.

    \section{\label{sec:results}Results}

    The phase-engineered fluorescence imaging system is used with a DH-PSF to localize single atoms in all three dimensions of the three-dimensional optical lattice.
    We present a scheme for calibrating the axial rotation angle of the DH-PSF. 
    As we show, this allows the vertical lattice structure to be resolved while preserving the lateral single-site resolution.

\begin{figure*}[t]	
    	\centering
        \includegraphics[width=\textwidth]{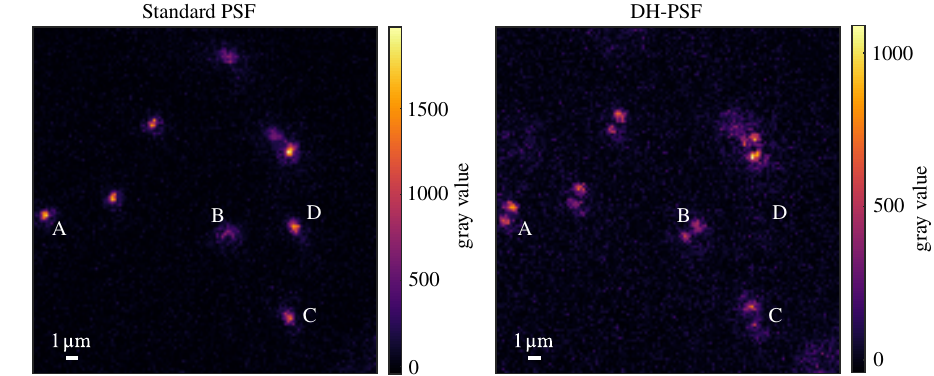}
    	\caption{Example of an image section taken with the standard PSF (left) and the DH-PSF (right).
        The (background corrected) images at $\NA=\num{0.6}$ with an exposure time of $\SI{1}{\second}$ have been consecutively taken from the same ensemble of atoms. Note that there is some atom hopping and loss between images. 
        Atom~A appears relatively in focus, while atom~B is clearly out-of-focus, resulting in a different angle of the DH-PSF relative to atom~A.  
        Atom~C also appears moderately defocused, but based on its DH-PSF angle it can be concluded that, unlike atom~B, it is located above the focal plane.
        Atom~D shows an example of an atom that is lost between images.
    	}
    	\label{fig:PSFcomparison}
    \end{figure*}

    Figure~\ref{fig:PSFcomparison} shows images of an axially thick atomic ensemble in a sparsely filled optical lattice taken with the DH-PSF compared to the standard PSF. 
    In the figure, atoms with a similar standard PSF but different lattice plane locations exhibit a clear difference in their DH-PSF. 
    That is, the standard PSF's symmetry about $z = 0$ is no longer present for the DH-PSF, allowing for one to discriminate between atoms on the positive or negative side of the focal plane. 
    As such, the DH-PSF contains more information about the atoms' position than the standard PSF. 
    Additionally, the DH-PSF enhances the depth of field of the imaging system; even atoms that look blurry with the standard PSF can be well localized with the DH-PSF. 
    Note, however, that the in-focus rotation angle of the DH-PSF is non-zero; we will return to this later in the manuscript.
    Finally, we observe a disadvantage of the DH-PSF---the energy is split between two peaks---giving a poorer signal-to-noise ratio per pixel compared to the standard PSF for an equal photon budget.

    To extract the atoms' position information from an image, one must determine the rotation angle of the DH-PSF as well as its center position.
    We low-pass filter the images (after a mean background subtraction) to limit the effect of shot noise.
    A local maxima search algorithm then finds the position of each peak.
    The pairwise separation of all peak combinations in an image is calculated. 
    Of these, only those pairs that are within a certain distance range, set by manual inspection of the detected PSFs, are retained.
    We then sort out those pairs where both peaks are simultaneously paired with each other. 
    This gives unambiguous pairs of peaks which, by construction, correspond each to a single atom.
    Finally, the position and rotation angle of the DH-PSF are determined by fitting a cropped region around the atom center (in the middle of the two peaks) to the sum of two elliptical two-dimensional Gaussian functions.

    The mapping from rotation angle to vertical position is, in theory, directly possible using Eq.~\eqref{eq:rotationangle}, as long as the Rayleigh length $\zR$ of the Laguerre-Gaussian mode superposition in object or image space is known.
    However, we define the $\zR$ of the mode combination by choosing the Laguerre-Gaussian waist $w_0$ of the phase mask in the pupil plane.
    More precisely, the resulting field in image space does not fully match the desired Laguerre-Gaussian mode superposition (cf.\ Sec.~\ref{sec:expt}), although the overlap is high. 
    Note that here the effective $w_0$ or $\zR$ defining the Laguerre-Gaussian basis with the highest overlap are unknown \textit{a priori}.
    A calibration of the vertical position is desirable to account for experimental effects not captured by this simple model (e.g., aberrations), even if the effective Rayleigh length $\zR$ is known.
    By shifting the focal plane by a known distance---by programming a holographic lens on the SLM---we obtain a vertical length reference which is used for calibration.
    The change in the measured angle of the image with shifted focus $\theta_s$ compared to the unshifted image $\theta$ is related to the displacement of the focal plane $\Delta z$ according to 
    \begin{equation}
        \label{eq:calibration}
        \theta_s = V\arctan{\bigg[ \tan{\bigg(\frac{\theta - \alpha}{V}\bigg)-\frac{\Delta z}{\zR}}\bigg]} + \alpha.
    \end{equation}

\begin{figure}[t]	
    	\centering
        \includegraphics[width=\columnwidth]{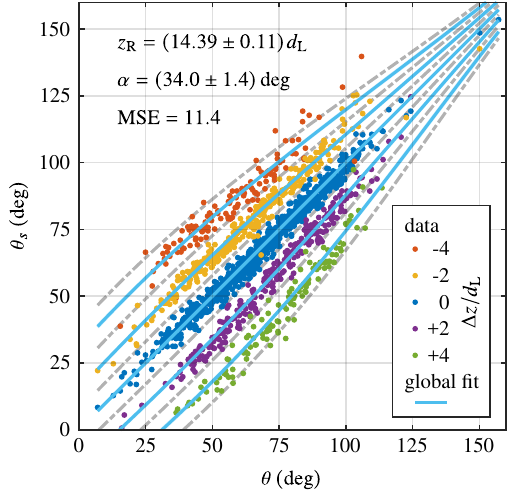}
    	\caption{Axial calibration of the DH-PSF giving the effective Rayleigh range $\zR$, the angle offset $\alpha$ and the mean squared error (MSE) of the nonlinear fit of Eq.~\eqref{eq:calibration} to the experimental data. The angle measured in the image with shifted focal plane $\theta_s$ is related to the angle measured in the unshifted images $\theta$ by Eq.~\eqref{eq:calibration}. Different focal plane shifts $\Delta z$ are shown in different colors. The lines corresponding to odd integer focal shifts, in units of the vertical lattice constant $\dz$, are marked in gray dashed lines.}
    	\label{fig:calibration}
    \end{figure}

    The experimental data sets with different shifts $\Delta z$ are (simultaneously) fitted to this model by non-linear least squares minimization.
    Here we take $V=2$ as defined by our DH-PSF.
    We post-select on atoms that have not hopped (in any Cartesian direction) around the lattice during imaging by requiring that $\theta_1 \approx \theta_3$, where we denote by $\theta_\ell$ the atom's rotation angle in image $\ell$, and both $\ell = 1,3$ are taken at the unshifted focus, and the approximation takes into account minor inaccuracies in the peak finding and angle determination algorithms
    as well as the systematic shift due to gravity, which will be discussed shortly.   
	The result is shown in Fig.~\ref{fig:calibration}.
    We can see that the measured angles follow Eq.~\eqref{eq:calibration} quite well.
    The in-focus angle $\alpha=\SI{34.0\pm1.4}{\deg}$ can be detected as a result of the nonlinear model.
    If the Eq.~\eqref{eq:calibration} was linear, $\theta_s$ would not depend on $\alpha$; this is clearly seen if we replace $\tan$ and $\arctan$ with the identity function.
    The effective Rayleigh length $\zR=(\num{14.39\pm 0.11})\,\dz$ is determined by the distance between stripes of equal $\Delta z$. This demonstrates the strongly extended depth of field of the DH-PSF as opposed to the theoretical $\lambda/(2\NA^2) \approx \num{2.2}\,\dz$ for standard PSF at $\NA=\num{0.6}$.

	Due to hopping along the optical axis of our system (here, the vertical direction, cf. Fig,~\ref{fig:expt_schematic}) during the exposure time, the data points are scattered around the fit lines corresponding to a particular, selected plane of the three-dimensional lattice, which is also evident from from the relatively large mean squared error of $\num{11.4}$, which is obtained from the nonlinear fit of Eq.~\eqref{eq:calibration} to the experimental data.
    We expect that a reduction in vertical hopping will result in measured angles lying closer to the relevant lines, 
    thus clearly resolving the different planes.
    Importantly, the fitted calibration parameters of our model in Eq.~\eqref{eq:calibration} are not affected by the atoms hopping between different planes.
    Such hopping events are not expected to produce a bias in the fitted parameters, as we explain in the following.
    Gravity makes it more probable for the atoms to hop vertically downwards during imaging, leading to a small bias in the DH-PSF angle.
    However, the fit includes both the $(\theta_1,\theta_2)$ and $(\theta_3,\theta_2)$ data pairs. 
    Their directional angle shifts due to gravity are statistically canceled out because they occur in opposite directions.
    This is not true for the data with $\Delta z=0$ for which we use the pairs $(\theta_1,\theta_3)$.
    Thus, in this case, gravity causes a systematic shift of the data points below the identity line $\theta_s = \theta$, which amounts in average to $\ang{0.63}$.
    Upon closer inspection, one can discern a higher density of blue points below the identity line in Fig.~\ref{fig:calibration}.
    However, this bias does not affect the calibration, since Eq.~\eqref{eq:calibration} yields the identity without any free parameters for $\Delta z=0$.
    As we will show, the accuracy of the axial length reference--given by the focus shift through the holographic lens--does not introduce any bias.

    The measured in-focus angle $\alpha$ deviates from the $\SI{0}{\deg}$ expected from the selected DH-PSF; we expect $\alpha = 0$ due to the fact that the SLM and camera axes coincide.
    This deviation arises due to residual aberrations. 
    As we will show in Sec.~\ref{sec:sims}, certain (low order) aberrations can have a strong influence on this offset angle.

    The calibration opens up the possibility of calculating the axial position of an atom by rearranging Eq.~\eqref{eq:rotationangle} into
    \begin{equation}
        z=\zR\tan\left(\frac{\theta-\alpha}{V}\right)\,.
    \end{equation}
    This can be employed, for instance, to determine the vertical distribution of atoms in the lattice.
    In this way, it is also possible to check the calibration against another known axial length reference, namely the distance between vertical lattice planes $\dz$.

\begin{figure}[t]
    	\centering
        \includegraphics[width=\columnwidth]{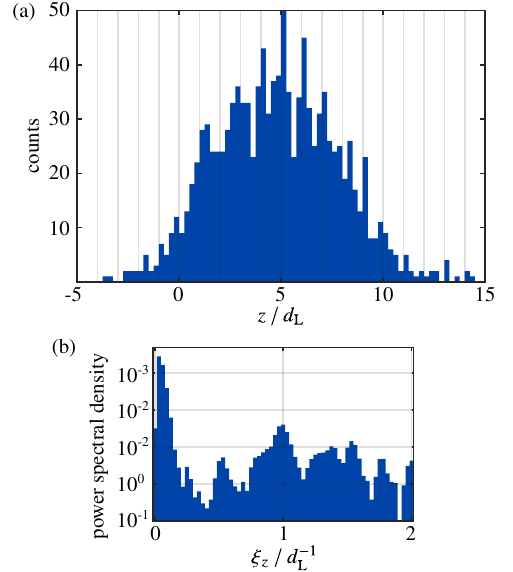}
    	\caption{ (a) Histograms of the calculated $z$ position of the atoms for a given focus setting, calculated from their measured angle $\theta$. (b) Fourier transform of the histogram in (a), showing a peak around $\xi_z = 1/\dz$, as expected due to the discrete nature of the lattice structure. This peak is, however, obscured due to between-planes hopping and inhomogeneous aberrations in the imaging system, as we describe in the main text.}
    	\label{fig:latticestructure}
    \end{figure}

    Fig.~\ref{fig:latticestructure}~(a) shows a histogram of the determined vertical positions of atoms imaged as described in Sec.~\ref{sec:expt}.
    For this purpose we take each of the first images of the dataset as for Fig.~\ref{fig:calibration}, a total of $\num{1173}$ atoms.
    Together with the determined lateral positions, the three-dimensional positions of the atoms in the lattice are thus reconstructed.
    The detected atoms lie mostly above the focal plane within a range of around ten vertical lattice planes centered around $z=\num{5}\,\dz$.
    This means that the focal plane of the microscope is shifted relative to the center of the atom cloud loaded into the lattice, which also shows that our method can be used as a means of calibrating the objective position relative to the position of the atoms loaded into the lattice.
    The enhanced depth of field by the DH-PSF is apparent from the fact that atoms can be readily detected more than $\num{\pm10}\,\dz$ away from the focus, i.e.\ a range of $\num{20}\,\dz$, as opposed to the $\num{2.2}\,\dz$ depth of field of the standard PSF. 

    A clustering at integer lattice sites is observed, as expected from the nature of the vertical lattice.
    To investigate the periodicity of the vertical atom distribution in more detail, we calculate the spatial frequency spectrum as shown in Fig.~\ref{fig:latticestructure}~(b) using the discrete Fourier transform.
    The spectrum peaks at the spatial frequency $\xi_z=\num{1}\,\dz^{-1}$, as expected.
    Vertical hopping during the image exposure leads to a smearing of the expected discrete structure (cf.\ the calibration fit shown in Fig.~\ref{fig:calibration}).
    Another contribution to the broadening arises due to aberrations inhomogeneously distributed over the field of view; in particular, the Strehl ratio of our objective lens decreases with increasing distance from the optical axis~\cite{alberti2017}.
    As such, in the following section, we present simulations that allow us to better understand the effect of the aberrations in the system.

\section{\label{sec:sims}Simulations}

    The basic idea of the simulations presented here is to physically model the light field emitted by a given atom (point source) in the pupil or SLM plane, and then to calculate the far field in the image or camera plane.
    The resulting intensity distribution corresponds to the PSF.
    PSFs for atoms at different axial positions are calculated by propagating the simulated image plane field for an in-focus atom along the optical axis. 
    The object plane is conjugate to the image plane and the respective $z$ coordinates are simply related via the axial magnification.

\begin{figure}[t]
    	\centering
        \includegraphics[width=\columnwidth]{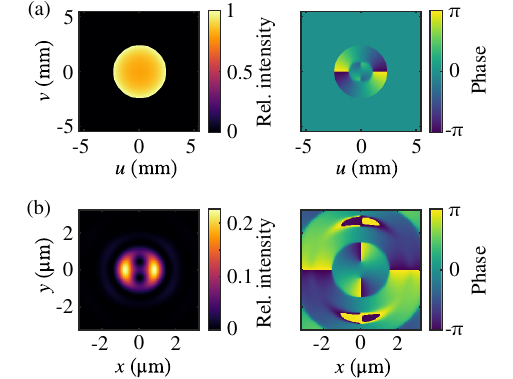}
    	\caption{Visualization of the simulated field emanating from a single atom without aberrations for a NA of $\num{0.6}$ (a) in the SLM plane and (b) in the image plane at $z=0$, after modification by the SLM into a DH-PSF. The SLM plane intensity is normalized to its maximum and increases towards the edges due to the apodization of the objective lens. The pupil radius corresponding to the NA of $\num{0.6}$ is $a=\num{230}$ SLM pixels and we use the same Laguerre-Gaussian waist $w_0$ as in the experiment such that $a/w_0=\num{2.97}$. Only a central section of the calculated output matrix is shown, so that the PSF is clearly visible. Here, the axes are given in object/atom plane coordinates. The image plane intensity is normalized to the maximal intensity of the standard PSF.}
    	\label{fig:simulationexample}
    \end{figure}

    The field in the pupil plane is given by the apodization function of the objective lens and the phase-modulated generalized pupil function, which captures arbitrary aberrations and the programmed DH-PSF phase mask.
    The pupil radius in the SLM plane $\NA \times f_2\times f_\mathrm{obj}/f_1$ corresponds, for $\NA=\num{0.6}$, to 230 SLM pixels, in our case.
    We define the pupil plane field by a $\num{1050}\times\num{1050}$ matrix with the physical pixel size corresponding to the SLM pixel size. 
    This selection is deemed reasonable, considering the smooth and gradual variation of the field resulting from the apodization and Zernike polynomials that characterize the aberrations across the pupil plane. 
    Phase discontinuities are introduced only by the phase mask programmed onto the SLM and therefore, due to the digital nature of the device, only occur at pixel boundaries.
    Intensity and phase are shown exemplarily without aberrations in Fig.~\ref{fig:simulationexample}~(a).

    We calculate the corresponding field in the image plane using the two-dimensional discrete Fourier transform.
    The output plane pixel size is given by $\lambda f_\text{tube}/d$ where $d$ is the physical side length of the pupil plane matrix.  
    Prior to the transformation, the matrix is subjected to zero-padding, increasing its size by a factor of 10, thus increasing the resolution in the output plane.
    The field at different axial positions in the image plane is computed by Fresnel propagation.
    The resulting transverse intensity distribution, shown in Fig.~\ref{fig:simulationexample}~(b) for $z=0$, corresponds to the SLM-modified DH-PSF; this allows the study of its inherent properties.

\begin{figure}[t]
    	\centering
        \includegraphics[width=\columnwidth]{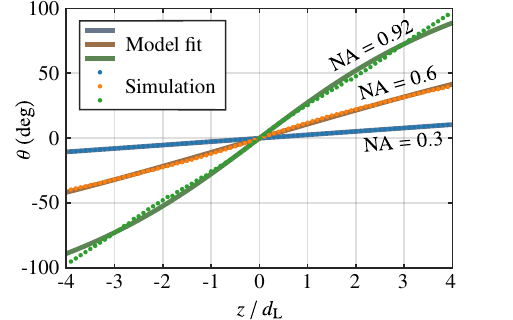}
    	\caption{Computed rotation angles and corresponding Laguerre-Gaussian model fits of the DH-PSF for three different NAs. The PSF was calculated at each of 81 axial positions in the interval from $\num{-4}$ to $\num{4}\,\dz$.}
    	\label{fig:simulationdifferntNAs}
    \end{figure}

    As an example, we can compute the DH-PSF's rotation angle for a range of axial positions, varying the NA of the system, as shown in Fig.~\ref{fig:simulationdifferntNAs}.
    The amount of rotation strongly depends on the NA; a high NA corresponds to a small radial scale $w_0$ and axial scale $\zR$, giving a faster rotation and defocus.
    A non-linear least squares minimization is used to fit the rotation angle to Eq.~\eqref{eq:rotationangle} describing the expectation for a perfect rendering of the Laguerre-Gaussian mode superposition. 
    We find that the smaller the NA, the better the fit of the Laguerre-Gaussian model. 
    The DH-PSF created solely by phase modulation is still quite well described by this model, in particular at low and intermediate NAs.\footnote{The SLM used in this work can only modulate phase; ideally a device would be used that could modulate both phase and amplitude.}
    Our method indeed creates the DH-PSF Laguerre-Gaussian mode superposition with high fidelity.
    Additionally, leakage into other Laguerre-Gaussian modes occurs symmetrically about the straight line in the $(l,p)$ Laguerre-Gaussian modal plane (cf.\ Sec.~\ref{sec:theory}), as was shown by Ref.~\cite{piestun2008a}, and thus has negligible influence on the rotational behavior.
    Hence, the result of these simulations supports our assumption that Eq.~\eqref{eq:rotationangle} represents a good estimator of the atoms' axial position based on the measured DH-PSF.
    In case of significant deviations, a look-up-table of an atom's axial position from the detected rotation angle can be established.

    Our simulations are further aimed at understanding which aberrations most affect the rotation angle of our DH-PSF. 
    From these, we find that the measured in-focus angle $\alpha$ is most sensitive to vertical astigmatism (both primary and secondary), as well as spherical aberration and, in a trivial way, defocus. 
    Thus, in the experiment, the degradation of the wavefront distortion away from the optical axis results in an inhomogeneous angle $\alpha$ across the field of view.
    Using an averaged angle $\alpha$, as determined by calibration, introduces an uncertainty in determining an atom's vertical position that is dependent on its lateral position within the field of view; this effect is not captured by our simplified model that does not include inhomogeneous aberrations. 
    Aggregating atoms across the entire field of view results in ``smearing'' of the measured angle at different focal positions; the peak of Fig.~\ref{fig:latticestructure}(b) is thus similarly smeared out.
    We know from previous work~\cite{alberti2017} that defocus and astigmatic effects are the main sources of aberration due to our objective, but other aberrations are likely due to other optical elements and misalignment of the system; these were only partially corrected by the SLM during the experiment.

    To study the effect of aberrations on the DH-PSF, we use the relation between the PSF and the wavefront. We express the aberrated pupil wavefront as an expansion of (lowest-order) orthogonal Zernike polynomials~\cite{tian2022}. 
    For our DH-PSF at the NA of 0.6, we simulate different types of aberrations by adding in the pupil plane each of the lowest-order non-trivial Zernike polynomials with varying Zernike coefficients. 
    We compute the resulting PSF in a range from $\num{-10}\,\dz$ to $\num{+10}\,\dz$ and determine the rotation angle.

    Since certain aberrations significantly change the double-peak structure of the DH-PSF above a certain strength, we determine the rotation angle not from a double Gaussian fit of the PSF intensity distribution as for the experimental data but rather using a nonparametric method based on the Radon transform.
    The Radon transform of the PSF contains the line projections of the intensity distribution under all projection angles and projection displacements. Here, the line projection refers to the sum of all pixel values in an image along a line with a given angle. For lines that are not horizontal or vertical, the algorithm used in this work chooses how to weight the individual pixel values in the sum.\footnote{We use the Radon transform algorithm implemented in the MATLAB function \texttt{radon} (version R2022b).} 
    We numerically compute the Radon transform of the simulated PSFs for projection angles in the interval $[-\pi/2,+\pi/2)$ with a resolution of $\SI{0.01}{\deg}$. 
    For a DH-PSF centered at the origin of the coordinate system, the rotation angle is the projection angle that corresponds to the maximum value of these Radon transforms at zero displacement.

\begin{figure*}[t]
    	\centering
        \includegraphics[width=\textwidth]{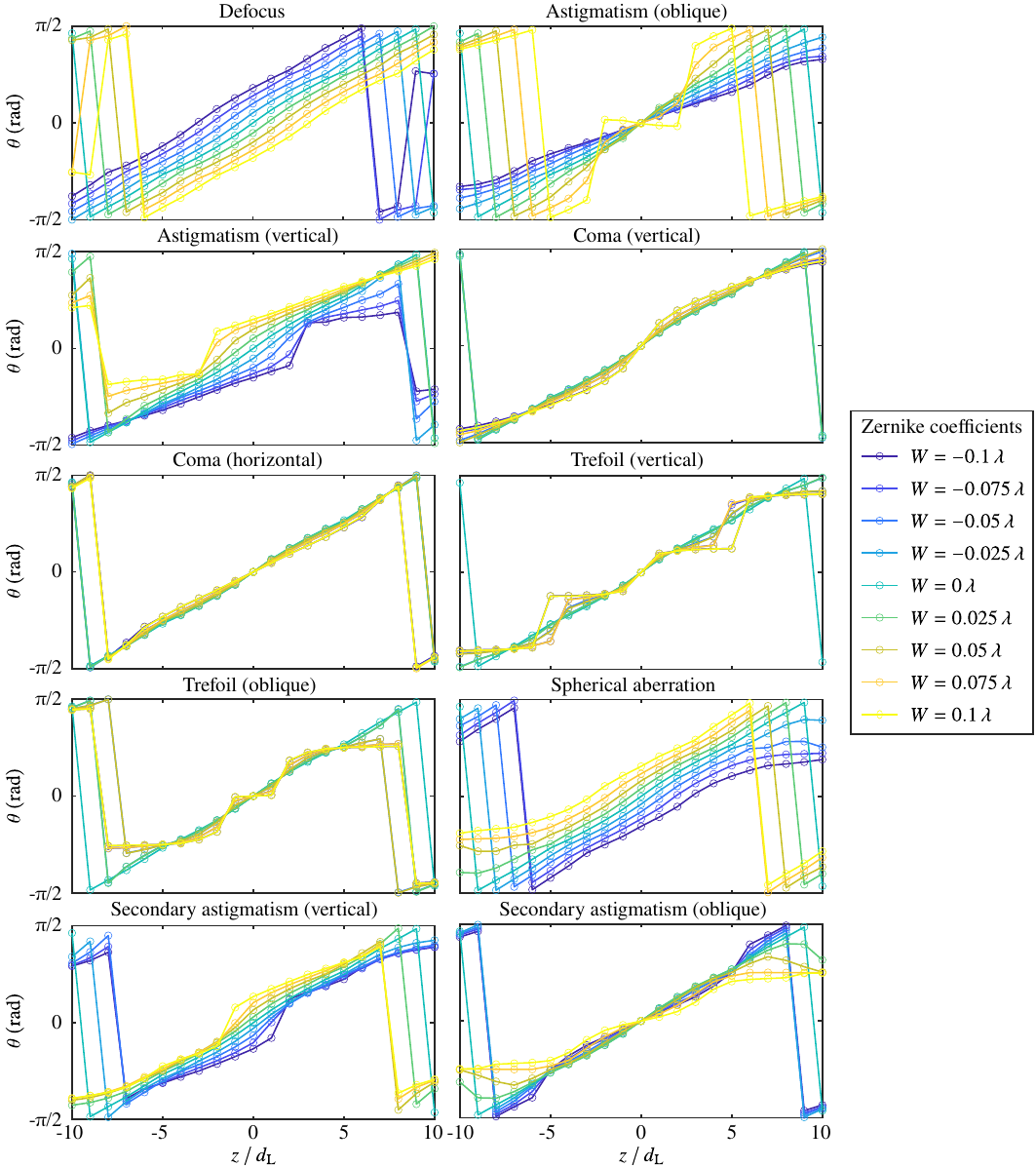}
    	\caption{Angle of rotation of the DH-PSF as a function of axial emitter position in the presence of aberrations for an NA of 0.6. 
     Each subplot shows the effect of a different Zernike polynomial wavefront aberration (up to Noll index 13, omitting the trivial piston, horizontal and vertical tilt) for varying Zernike coefficients or aberration strengths $W$. 
     The rotation angle $\theta$ is determined from the computed DH-PSF using the Radon transform for every integer position within the axial range of $\num{-10}$ to $\num{10}\,\dz$ (see text for details). 
     The rotation angle of our DH-PSF with $V=2$ is only unique within an angle interval of length $\pi$. 
     Since our angle determination is restricted to $\theta\in[-\pi/2,+\pi/2)$ by construction, the computed angles jump by $\pm\pi$ at the edges of this interval.
     }
    	\label{fig:simulation_aberration}
    \end{figure*}

    Fig.~\ref{fig:simulation_aberration} shows the results of this analysis.
    We find that the different types of aberrations affect the in-focus rotation angle $\alpha = \theta_{z=0}$ to very different extents.
    As expected, the defocus trivially shifts the entire rotation curve along $z$, thereby also impacting $\alpha$.
    From the other aberrations, only vertical astigmatism (both primary and secondary) and spherical aberration have a substantial impact on $\alpha$, all to a similar extent.
    Unsurprisingly, the simulation reveals that a higher absolute value of the Zernike coefficient (corresponding to a stronger aberration) results in a greater shift in the rotation angle.
    As can be seen, the rotation angle is also affected at $z\ne 0$, especially for oblique and vertical primary astigmatism, spherical aberration, and vertical secondary astigmatism.
    The impact of spherical aberration closely resembles that of defocus but with an opposing sign.
    While vertical and horizontal coma, and to some extent vertical and oblique trefoil as well as oblique secondary astigmatism do not significantly alter the rotation curves, we find that these aberrations do change the shape of the DH-PSF at higher Zernike coefficients.

    While we do not know the exact residual aberrations present in our imaging system, the simulation of the DH-PSF with aberrations presented here leads us to conclude that certain aberrations have an appreciable influence on the rotational behavior of the DH-PSF.
    In addition to defocus, vertical astigmatism (primary and secondary) and spherical aberration shift the rotation curves.
    It is therefore plausible that the measured in-focus angle $\alpha$ deviates significantly from zero due to such aberrations (cf.~Sec.~\ref{sec:results}).
    Since we know that the aberrations are not completely homogeneous across the field of view, we must also conclude that the calibration of the rotation angle, which is based on an aggregation of PSFs from different lateral positions, leads to a value of $\alpha$ averaged over our field of view. 
    This leads to a small uncertainty in the angle determination depending on the lateral atom position. 
    Hence, optical aberrations, together with between-planes hopping, are believed to be the main reason for the imperfect separation of different lattice planes, as observed in Fig.~\ref{fig:calibration} and~\ref{fig:latticestructure}.

\section{\label{sec:conc} Discussion and Conclusion}

    This work describes a method to accurately determine an atom's position in a quantum gas microscope in three-dimensional space. 
    The data presented is currently limited by aberrations in the system, but the lattice structure that is nevertheless visible in Fig.~\ref{fig:latticestructure} demonstrates that the rigidly-rotating, double-lobed DH-PSF has the capability of resolving single planes of an optical lattice with typical sub-micron spacing. 
    Additionally, the engineered PSF greatly increases the depth of field compared to a typical Airy-like PSF. 
    Our data furthermore confirms the accuracy of the axial length reference used for the calibration of axial position detection, i.e., the use of a focal shift by a holographic lens on the SLM. This use of a holographic lens for shifting the focal plane of our atom images is a much simpler alternative to the common technique of piezo control of the objective in systems in which an SLM is used for PSF engineering~\cite{moerner2009} or systems involving spatial atomic tomography with multiple imaging exposures~\cite{sherson2020}. 

    Our simulations demonstrate how aberrations impact the rotation angle of the DH-PSF and to what extent. 
    Specifically, our simulations show how the rotation angle of an aberrated DH-PSF in the focal plane depends on the type and degree of aberrations present in the imaging system. 
    Inhomogeneities of aberrations across the field of view thus introduce an uncertainty in position determination.
    This effect was also observed and characterized in DH-PSF microscopy of single molecules~\cite{moerner2015} where, to calibrate the system, the PSF was sampled across the field of view with an accurate axial and lateral length reference, using a regularly spaced sub-diffraction aperture grid filled with fluorescent dyes. 
    Effectively, the simple calibration method presented here for extracting the axial position from the detected PSF's rotation angle averages over atoms detected within a larger region of the field of view. 
    Due to the regular geometry of our lattice, such a calibration parameterized over the field of view is possible in the future using experimental data of the type we recorded. 
    By mapping aberrations over the field of view, it should be possible to mitigate uncertainties caused by inhomogeneous aberrations and, thereby, enhance the precision of axial localization. However, we expect that in microscopes with better aberration homogeneity, the proposed mapping of aberrations over the field of view will not be necessary to resolve typical optical lattice spacings over the entire field of view of interest.
    Since the DH-PSF is sensitive to (certain) aberrations, the modified PSF could also be used for precise aberration characterization. Moreover, we remark that the SLM in the intermediate Fourier plane then allows the addition of a phase map that compensates for the detected aberrations, regardless of whether a DH-PSF or a regular Airy-like PSF is used; another method of aberration correction can be found in Ref.~\cite{spielman2021}.

    For quantum gas microscope experiments with densely filled optical lattices~\cite{alberti2017a}, it is of interest not to lose the lateral single-site resolution due to the widening of a modified PSF, leading to an overlap in the signals from nearby atoms.
    Taking the DH-PSF as an example, the larger extent of the PSF and the task of correctly attributing lobes of overlapping PSFs to the right atoms can initially be believed to be an inherent disadvantage. 
    However, as reported in Ref.~\cite{piestun2014}, the particular shape and spacing of the lobes can help determine the underlying positions of the emitters. 
    In fact, similar to the one-dimensional PSF case described in Ref.~\cite{meschede2016}, relatively dense ensembles in which the PSFs of close-by emitters overlap can be super-resolved, in all three dimensions~\cite{piestun2014}.
    This is supported, to some extent, by the Fisher information calculations presented in Fig.~\ref{fig:dhpsf_fisherinformation}, although further work remains to be done along these lines. For example, a complete noise model would allow the theoretical resolution of the system to be calculated \cite{Ober:2004}.
    In particular, one can explore non-helicoidal PSFs that can provide sufficient information about an atom's $(x,y,z)$ position without breaking into two rotating points, which could make the application of deconvolution and other postprocessing algorithms (as in Ref.~\cite{kuhr2022}) much easier.
    Further optimizations of rotating PSFs have been reported, such as iteratively optimizing the phase pattern programmed onto the SLM by enhancing the Laguerre-Gaussian modal components that define exact PSF rotation~\cite{piestun2008a}, or maximizing the Fisher information of the PSF by varying the number and position of vortex singularities using an analytical expression for the phase mask~\cite{piestun2012}. 
    Yet another approach maximizes Fisher information while restricting the phase degrees of freedom to the lower Zernike modes~\cite{moerner2014}.
    These improvements will possibly allow for the accurate three-dimensional reconstruction of an atomic distribution within a quantum gas microscope using a single experimental exposure, even in the regime of high filling of the lattice sites. Furthermore, we would like to highlight that our methods are not limited to quantum gas microscopes; any atomic or ionic system that provides a fluorescence image, e.g., trapped ions or atoms in optical tweezers, could utilize similar methods. The work presented here thus provides the proof-of-principle for future work that extends the power of such quantum simulators into studies of more complex three-dimensional systems. 

\section{Acknowledgments}
    This research was supported by the Collaborative Research Center SFB/TR 185 OSCAR of the German Research Foundation. CW acknowledges support from the University of Bristol Schools of Physics and Electrical, Electronic, and Mechanical Engineering. 
    We thank B.~Bernard for contributions in the early stage of the presented work and E.~Uruñuela for fruitful discussions.

\appendix{\label{app:app}}

\section{Superposition of Laguerre-Gaussian modes}\label{app:lgsuperpositions}
    We present here derivations concerning the rotating PSFs from the superposition of Laguerre-Gaussian modes. 
    This derivation is analogous to the one in Ref.~\cite{shamir1996}, but due to the different notation and choice of normalization, we present it here for the sake of clarity.  

    The Laguerre-Gaussian transverse modes of order $(l,p)$ (orthogonal basis set of solutions to the paraxial Helmholtz equation in cylindrical coordinates) are given by
    \begin{multline}\label{eq:appendixLGmodes}
        u_{lp}(r,\phi,z) = C_{lp} \,
        \underbrace{\frac{w_0}{w(\tilde{z})}
        \exp\left(-\tilde{r}^2\right)
        \exp\left( \i \tilde{r}^2\tilde{z}\right)
        }_{=:\, G(r,z)} \\
        \times \underbrace{
        \left(\sqrt{2}\tilde{r}\right)^{|l|}
        L_p^{|l|}\left(2\tilde{r}^2\right) 
        }_{=:\, R_{lp}(r,z)}
        \exp\left(\i l\phi -\i\psi_{lp}(\tilde{z})\right)\,,
    \end{multline}
	with the definitions given in Sec.~\ref{sec:theory}. 
    The field $\vect{U}_{lp}(\vect{x},t)$, for clarity, is then given by the components $U_{lp}(\vect{x},t) = \tilde{U}_{lp}(\vect{x})\, \e^{-\i\omega t}$ with $\tilde{U}_{lp}(\vect{x}) = u_{lp}(r,\phi,z)\, \e^{\i kz}$.
    The paraxial Helmholtz equation
	\begin{equation}\label{eq:paraxialhelmholtzequation}
		\nabla_\perp^2 \tilde{U}_{lp} + 2\i k \diffp{\tilde{U}_{lp}}{z} = 0
	\end{equation}
    has the same form as the free space Schrödinger equation in two dimensions under substitution of the $z$ coordinate by the time $t$, the transverse mode function 
	$u_{lp}(\vect{x})=\braket{\vect{x}|l,p}$ by the wave function $\Psi(x,y,t)=\braket{x,y\,|\,\Psi(t)}$ and the wave number $k$ by $m/\hbar$.
	This allows us to use the quantum mechanics formalism for the analysis of paraxial waves by using time domain semantics to describe the evolution along the $z$-axis, e.g.\ the convenient use of Dirac notation.
	For instance, the orthonormality relation of the Laguerre-Gaussian modes can concisely be written as
	\begin{equation}
		\braket{l',p'|l,p} = \delta_{ll'}\delta_{pp'}\,,
	\end{equation}
	with the standard scalar product.

	Let us examine a superposition of $N$ Laguerre-Gaussian modes with normalized coefficients $a_j\in\mathbb{C}$,
	\begin{equation}
		\ket{A} = \sum_{j=1}^{N} a_j \ket{l_j,p_j}\,,\quad\text{where}\quad \sum_{j=1}^{N} \abs{a_j}^2 = 1
	\end{equation}
	with the modes sorted according to their combined mode indices $n_j = 2p_j + \abs{l_j}$ such that $n_j\le n_{j+1}$.
	The intensity is (using the abbreviations defined in 
	Eq.~\eqref{eq:appendixLGmodes}) then given by
	\begingroup
	\allowdisplaybreaks
	\begin{alignat}{3}
		I(\vect{x}) &=&&\, \frac{\epsilon_0 \cvacuum}{2} \left|\braket{\vect{x}|A}\right|^2 
		= \frac{\epsilon_0 \cvacuum}{2}\, \left|\,\sum_{j=1}^{N}a_j\braket{\vect{x}|l_j,p_j}\,\right|^2 \\
		&=&&\, \frac{\epsilon_0 \cvacuum}{2}\, \sum_{j=1}^{N}\sum_{k=1}^{N} a_j 
		a_k^*\braket{\vect{x}|l_j,p_j}\braket{l_k,p_k|\vect{x}} \\
		&=&& \,\frac{\epsilon_0 \cvacuum}{2}
		\left( \sum_{j=1}^{N}  \abs{a_j}^2 \left|\braket{\vect{x}|l_j,p_j}\right|^2 \right. \\
		&&&\quad\,+  2\sum_{j=1}^{N}\sum_{k=j+1}^{N}
		\left.\Re\left(a_j a_k^* \braket{\vect{x}|l_j,p_j} \braket{l_k,p_k|\vect{x}}\right) \vphantom{\sum_{j=1}^{N}} \right) \nonumber\\
		&=&& \,\frac{\epsilon_0 \cvacuum}{2}
		\left( \sum_{j=1}^{N}  \abs{a_j}^2 \left|\braket{\vect{x}|l_j,p_j}\right|^2 \right.  \\
		&&& \quad\,+  2\sum_{j=1}^{N}\sum_{k=j+1}^{N}
		\left(\abs{a_j} \abs{a_k} \left|\braket{\vect{x}|l_j,p_j}\right| 
		\left|\braket{\vect{x}|l_k,p_k}\right|\right) \nonumber\\
		&&&\quad\times\cos\left( \arg(\braket{\vect{x}|l_j,p_j}) - \arg(\braket{\vect{x}|l_k,p_k}) \right. \nonumber \\ 
		&&& \quad+ \left.\left. \arg(a_j) - \arg(a_k) \right) \vphantom{\sum_{j=1}^{N}}\!\right) \nonumber\\
		&=&& \,\frac{\epsilon_0 \cvacuum}{2}\, \abs{G(r,z)}^2\,
		\left( \sum_{j=1}^{N}  \abs{a_j}^2 C_{l_jp_j}^2 R_{l_jp_j}^2(r,z) \right.  \label{eq:intensitysuperposition}\\
		&&& \quad+ 2\sum_{j=1}^{N}\sum_{k=j+1}^{N}
		\abs{a_j}\abs{a_k}C_{l_jp_j}C_{l_kp_k}
		R_{l_jp_j} R_{l_kp_k} \nonumber\\
		&&&\quad\times\cos\left((l_j-l_k)\phi - (n_j-n_k)\arctan(z/\zR)\right. \nonumber \\
        &&&\quad + \left.\left. \arg(a_j)-\arg(a_k)\right) \vphantom{\sum_{j=1}^{N}}\!\right)\,. \nonumber 
	\end{alignat}
	\endgroup
	The first sum is axially symmetric and stationary in $z$ except for a scaling with $w(z)$.
	Therefore, these terms do not contribute to the targeted scaled-rigid rotation.
	The terms in the second sum rotate linearly with $\arctan(z/\zR)$ at the rotation rates
	\begin{equation}\label{eq:rotationvelocityonemodepair}
		\left(\diff{\phi}{z}\right)_{jk} = \frac{\Delta n_{jk}}{\Delta l_{jk}} \diff{}{z}\arctan(z/\zR)\,,
	\end{equation}
	where $	\Delta n_{jk} := n_j-n_k$ and $\Delta l_{jk} := l_j-l_k$.

	As can be seen in the cosine term of Eq.~\eqref{eq:intensitysuperposition}---in order to break the axial symmetry---at least two modes must have different azimuthal mode numbers, $\Delta l_{jk} \ne 0$, such that their different orbital angular momentum phases $\e^{\i l_j \phi}$ and $\e^{\i l_k \phi}$ give an azimuthal interference pattern.
	Similarly, the necessary condition for dynamic behavior along the $z$-axis is that  $\Delta n_{jk} \ne 0$ for at least two modes.
	The different Gouy phases $\psi_{l_jp_j}$ and $\psi_{l_kp_k}$ then lead to $z$-dependent interference patterns.

	Scaled-rigid rotation occurs if and only if
	all interference terms rotate at the same velocity.
	Evidently, a \emph{necessary} condition for this is that
	\begin{equation}
		\frac{n_{j+1}-n_j}{l_{j+1}-l_j} =: \frac{\Delta n_j}{\Delta l_j} =: V_j
	\end{equation}
	is the same for all Laguerre-Gaussian modes, that is, 
	\begin{equation}\label{eq:appendixconditionscaledrigidrotation}
		V:=V_j=\text{const.} \quad\forall j\in\{1,2,\dots, N\}\,.
	\end{equation}
	This condition is also \emph{sufficient}, as can easily be by proven by induction:
	Assume that Eq.~\eqref{eq:appendixconditionscaledrigidrotation} holds. As by definition $n_j = 2p_j+\abs{l_j}$, it 
	follows that both 
	$\Delta n_j = \text{const.} =:\Delta n$ and $\Delta l_j = \text{const.}  =:\Delta l$ hold. 
	We now need to show that $\forall k 
	\in\{1,2,\dots,N\}$ with $k\ne j$ the fraction of mode number differences are equal to $V$, i.e.\
	\begin{equation}\label{eq:inductionproof}
		\frac{n_k-n_j}{l_k-l_j} = V\,.
	\end{equation}
	Without loss of generality, we can take $k > j$ (if $k<j$, just rename $k\leftrightarrow j$).
	From our assumption directly follows that Eq.~\eqref{eq:inductionproof} holds for $k=j+1$. 
	Thereby also $n_k-n_j=\Delta n$ and $l_k-l_j=\Delta l$ separately hold for $k=j+1$.
	We proceed by showing the induction step $k\rightarrow k+1$
	\begin{equation}
    \begin{split}
		\frac{n_{k+1}-n_j}{l_{k+1}-l_j} &= \frac{n_k-n_j+n_{k+1}-n_k}{l_k-l_j+l_{k+1}-l_k} 
		\\ & = \frac{\Delta n + \Delta n_k}{\Delta l + \Delta l_k} = \frac{2\Delta n}{2\Delta l} = V
    \end{split}
	\end{equation}
	by using the induction hypothesis. It follows that Eq.~\eqref{eq:appendixconditionscaledrigidrotation} is also a 
	sufficient condition for scaled-rigid rotation.

	In particular, any combination of only two Laguerre-Gaussian modes yield scaled-rigid rotation.
	We can conclude that for mode combinations meeting the condition of scaled-rigid rotation, the rotation velocity of 
	the transverse intensity pattern can be written according to Eq.~\eqref{eq:rotationvelocityonemodepair} as
	\begin{equation}
		\diff{\phi}{z}= V \diff{}{z}\arctan(z/\zR)\,,
	\end{equation}
	which is nothing else but Eq.~\eqref{eq:phi_rot}. 
    Integration gives the rotation angle of the intensity pattern, i.e.\ Eq.~\eqref{eq:rotationangle}.

\section{Calculation of the Fisher information}\label{app:fisherinformation}
    We present the details on the computation of the Fisher information for the Laguerre-Gaussian fundamental mode and the DH-PSF. 
	Engineering a rotating PSF only makes sense if it also provides better three-dimensional position information than the existing regular PSF.
    The information theoretic approach mentioned in Sec.~\ref{sec:theory} provides the mathematical framework to study this question and is introduced e.g.\ in Ref.~\cite{kay1993}.

    To define the Fisher information, consider an observable $X$ whose probability depends on a parameter $\eta$.
    The Fisher information describes the amount of information that the observable $X$ contains about the unknown parameter $\eta$.
    For a single parameter $\eta$ and the probability density function $f(X;\eta)$, the Fisher information is defined to be the variance of the partial derivative of the log-likelihood function with respect to the parameter
    \begin{alignat}{3}
        \label{eq:fisherinformation2}
        \mathcal{I}_\eta &=&& \operatorname{E}\left[ \left.\left(\diffp{}{\eta}\ln f(X;\eta)\right)^2\,\right|\eta \right] \nonumber\\
        &=&& \int \left(\diffp{}{\eta}\ln f(\chi;\eta)\right)^2 f(\chi;\eta) \dl{\chi}\,.
    \end{alignat}
    Note that it does not have to be a probability distribution over a one-dimensional space. 
    The distribution is assumed to be single-parametric here; for multi-parameter models, the Fisher information can be written as a matrix.

    A normalized transverse intensity distribution can be understood as a probability density. 
    Thus, Eq.~\eqref{eq:fisherinformation2} can be used to calculate and compare the Fisher information with respect to a spatial coordinate $\eta\in\{x,y,z\}$ for transverse intensity distributions 
    \begin{equation}\label{eq:transverseintensitydistribution}
        f(x,y;\eta) = \left|\left. u\right|_{z=z_0}\right|^2
    \end{equation}
    at axial position $z_0$.
    Here, $u$ is the normalized transverse mode as defined in Appendix~\ref{app:lgsuperpositions}.
    Accordingly, the integration $\dl{\chi}$ is along both transversal dimensions.
    For the fundamental Laguerre-Gaussian mode, one obtains the simple analytical expressions
    \begin{alignat}{3}\label{eq:fisherinformationoffundamentalmode}
        \mathcal{I}_z &=&&\, 4\left(\frac{w_0}{\zR}\right)^2\frac{\tilde{z}^2}{\left(1+\tilde{z}^2\right)^2}\qquad \text{and}\\
        \mathcal{I}_x&=&&\,\mathcal{I}_y = \frac{4}{1+\tilde{z}^2}\,.
    \end{alignat}
    For more complex expressions, such as those arising in the DH-PSF $\ket{l,p}=\left(\ket{0,0}+\ket{2,1}\right)/\sqrt{2}$, we compute the derivative and integration numerically. 
    To evaluate the numerical error introduced hereby, the numerical calculation for the fundamental mode was compared with the analytical result from Eq.~\eqref{eq:fisherinformationoffundamentalmode}, revealing very good agreement.

    Here, for simplicity, we assumed no noise or other detrimental effects (such as a limited mode fidelity) so that the Fisher information can be calculated analytically from the theoretical intensities.
    However, when considering a limited signal-to-noise ratio and taking into account the limited mode fidelity, the DH-PSF provides higher and more uniform Fisher information for three-dimensional localization as compared to using a standard PSF, as has been shown in Refs.~\cite{piestun2008,piestun2009,harvey2019}.  

\section{Focal plane shift of the holographic lens}\label{app:holographiclens}
    We show the calculation of the displacement of the focal plane introduced by a holographic lens displayed on the SLM in the setup described in Sec.~\ref{sec:expt} (cf.\ Fig.~\ref{fig:expt_schematic}).
    A quadratic phase
			\begin{equation}\label{eq:quadraticphase}
				\varphi_\text{hol}(u,v) = - k\frac{u^2+v^2}{2 f_\text{hol}} \mod 2\pi\,,
			\end{equation}
	where $k=2\pi/\lambda$, is programmed on the SLM.
    This can be understood as a controlled insertion of a defocus aberration, or equivalently as programming a holographic lens or Fresnel lens of focal length $f_\text{hol}$ located in the SLM plane.
	Let us first look only at the lens system consisting of holographic lens and the lens with $f_2$ creating the intermediate image.
	The distance between these shall be $d$. 
	The shift is calculated from the difference of the front focal length of the lens system with and without the holographic lens
	\begin{equation}
		\Delta z_\text{I} = f_2 - \mathrm{FFL} = f_2 - \frac{f_2(d - f_\text{hol})}{d-(f_2+f_\text{hol})} 
	\end{equation}
	which for $d=f_2$ yields
	\begin{equation}
		\Delta z_\text{I} =  \frac{f_2^2}{f_\text{hol}}\,.
	\end{equation}
	The corresponding shift $\Delta z$ with respect to the initial focal plane at the atoms is then related to $\Delta z_\text{I}$ by the axial magnification $f_\text{obj}^2/f_1^2$ and yields
	\begin{equation}
		\Delta z = \frac{f_\text{obj}^2 f_2^2}{f_1^2 f_\text{hol}}\,.
	\end{equation}

\newpage

\end{document}